\documentclass[12pt]{article}

\begin{document}

\thispagestyle{empty} \noindent {\small \hfill hep-th/0211288 }

{\vskip2.0cm}

\begin{center}
{\Large \textbf{AdS$_{2}$ D-Branes in Lorentzian AdS$_{3}$}}

{\vskip0.5cm}

\textbf{Cemsinan Deliduman}

{\vskip0.5cm}

\textit{Feza G\"{u}rsey Institute}

\textit{Emek Mahallesi, No: 68}

\textit{\c{C}engelk\"{o}y 81220, \.{I}stanbul, Turkey}

\texttt{cemsinan@gursey.gov.tr}

{\vskip1.0cm}

\textbf{Abstract}

{\vskip0.5cm}
\end{center}

The boundary states for AdS$_{2}$ D-branes in Lorentzian AdS$_{3}$
space-time are presented. AdS$_{2}$ D-branes are algebraically defined by
twisted Dirichlet boundary conditions and are located on twisted conjugacy
classes of SL(2,R). Using free field representation of symmetry currents in
the SL(2,R) WZNW model, the twisted Dirichlet gluing conditions among
currents are translated to matching conditions among free fields and then to
boundary conditions among the modes of free fields. The Ishibashi states are
written as coherent states on AdS$_{3}$ in the free field formalism and it
is shown that twisted Dirichlet boundary conditions are satisfied on them.
The tree-level amplitude of propagation of closed strings between two AdS$%
_{2}$ D-branes is evaluated and by comparing which to the characters of $%
\widehat{\mathtt{sl}}$(2,R) Kac-Moody algebra it is shown that only states
in the principal continuous series representation of $\widehat{\mathtt{sl}}$%
(2,R) Kac-Moody algebra contributes to the amplitude and thus they are the
only ones that couple to AdS$_{2}$ D-branes. The form of the character of $%
\widehat{\mathtt{sl}}$(2,R) principal continuous series and the boundary
condition among the zero modes are used to determine the physical boundary
states for AdS$_{2}$ D-branes.

\newpage \pagenumbering{arabic}

\section{Introduction}

Three dimensional Anti-de Sitter space-time (AdS$_{3}$) is the most
accessible example in the quest to understand properties of string theory on
curved space-times with curved time. AdS$_{3}$ is the covering space of the
non-compact SL(2,R) group manifold and hence one can employ SL(2,R)
Wess-Zumino-Novikov-Witten (WZNW) model in order to formulate string theory
on AdS$_{3}$. There has been much work \cite{BRFW}--\cite{MO-1} on string
theory on SL(2,R) group manifold or on its covering space AdS$_{3}$. The
motivation to understand the string motion in the background of two
dimensional SL(2,R)/R black hole \cite{BN},\cite{witten} and three
dimensional BTZ black hole \cite{BTZ} also contributed to the motivation to
study SL(2,R)\ WZNW\ model.

From the start SL(2,R) WZNW\ model presented several new challenges unseen
in the formulation of string theory in flat space-time. The most important
of these is the unitarity problem. In flat space-time the negative norm
states due to time dimension can be eliminated successfully by using
Virasoro constraints and a no ghost theorem can be proved \cite{thorn}.
However, in SL(2,R)\ WZNW\ model it is found that even after imposing the
Virasoro constraints some negative norm states still remain in the spectrum.
To resolve this problem of unitarity and to get rid of ghosts two main
proposal are made. 1) The eigenvalue of the compact generator of SL(2,R), $j$%
, and the WZNW level, $k$, need to be restricted \cite{BN},\cite{others1} as
in the case of SL(2,R)/U(1) coset model \cite{peskin}. 2) The effect of zero
modes of free fields in the free field formulation of SL(2,R)\ WZNW\ model
should be carefully taken into account \cite{bars1}. According to the first
proposal, the closed string states belong to the highest and lowest weight
discrete as well as principal continuous series representations of $\widehat{%
\mathtt{sl}}$(2,R)\footnote{%
Notation: SL(2,R) stands for the group, \texttt{sl}(2,R) for the Lie algebra
and $\widehat{\mathtt{sl}}$(2,R) for the Kac-Moody algebra.}. However, due
to the mentioned extra conditions the closed strings in highest and lowest
weight discrete series representations cannot be excited above the first
excited level. Also only the lowest level states are allowed form the
principal continuous series representation. In \cite{MO-1} this proposal is
developed by also taking into account the contribution of possible CFT on
direct product spaces so as to make the total central charge equal to the
critical value 26. Then, considering also the effect of spectral flow, one
finds a finite number of excited states. However if the contribution of
other CFT is taken to be zero, i.e. taking SL(2,R) WZNW model by itself
critical, formalism gives very a few excited string states \cite{larsen}.
These arguments and being the restrictions on $j$ and $k$ artificial makes
this proposal not very attractive to us. The second proposal requires the
closed string states to belong only to the principal continuous series
representation of $\widehat{\mathtt{sl}}$(2,R) supplemented with zero modes.
In this spectrum there are no negative norm states and the string can be
excited to any arbitrary level \cite{bars1}. In \cite{BDM} this idea is
further supported by showing that the vertex operators for the lowest level
closed string states, written for finite value of WZNW level $k$, indeed
approach to the correct vertex operators in the flat $\mathcal{M}_{3}$
space-time, obtained as the limit of AdS$_{3}$ space-time as $k\rightarrow
\infty $. Mass-shell condition in AdS$_{3}$ space-time is also shown to
become the correct mass-shell condition in flat space-time in that limit.
The presence of all the zero modes is important in getting the correct
mass-shell condition in the flat limit. Setting any of the zero modes to
zero results in an incorrect expression.

The conclusions of \cite{bars1},\cite{BDM} is supported in the present paper
by considering the coupling of closed strings to AdS$_{2}$ D-branes in
Lorentzian AdS$_{3}$. D-branes in flat string theory are defined through
Dirichlet conditions at the open string end-points and they carry
non-vanishing vector field on their world volume which couples to those open
string end-points \cite{TASI}. Formulation of Dirichlet and Neumann boundary
conditions for the open string theories on general group manifolds has been
done in \cite{stanciu-0} by using the analogy between Kac-Moody symmetry
currents of WZNW model and the primary fields of the string theory in flat
space-time. The analogs of Dirichlet and Neumann conditions among Kac-Moody
symmetry currents are called as gluing conditions and from the
representation of currents in terms of group element it is found that
classically D-branes are located on the regular or twisted conjugacy classes
of that group \cite{stanciu-0}. The specific classical theory of D-branes on
non-compact SL(2,R) group manifold is discussed in \cite{stanciu},\cite
{bachas}. In the case of SL(2,R) WZNW model the possible gluing conditions
are checked in \cite{lomholt} to find out if they can be obtained from the
variation of the WZNW action. We review this analysis in section 2.2.
Quantum theory of AdS$_{2}$ and other D-branes in Euclidean AdS$_{3}$
(SL(2,C)/SU(2) coset WZNW model) are discussed by several authors \cite
{ads2d}--\cite{ponsot}. Different proposals for one-point functions and
boundary states for AdS$_{2}$ D-branes in Euclidean AdS$_{3}$ space were
presented in these works.

It has been hoped that the theory of D-branes in Lorentzian AdS$_{3}$ can be
obtained by performing Wick rotation in the theory in Euclidean AdS$_{3}$
\cite{ads2d}--\cite{ponsot}. However, we find it safer to formulate the
theory in Lorentzian AdS$_{3}$ without making a Wick rotation at the very
beginning. One of the reasons of this is that as it is discussed in \cite
{roberts} the Wick rotation in a curved space-time is more elaborate and it
might not work as in the case of flat space-time. Other than this, we note
that an important difference between Euclidean and Lorentzian theories were
already pointed out in the closed string theory case in respective
backgrounds. The vertex operators written in the Lorentzian AdS$_{3}$ in
SL(2,R)\ WZNW model \cite{BDM} contains an extra non-trivial factor compared
to the ones written in the Euclidean AdS$_{3}$ in SL(2,C)/SU(2)\ WZNW model
\cite{teschner-1}. Similarly we claim that the elements (boundary states,
correlation functions, etc.) of the theory of D-branes in Euclidean AdS$_{3}$
may not be, after Wick rotation, the correct elements of the theory of
D-branes in Lorentzian AdS$_{3}$. Therefore our strategy is to formulate the
theory first in Lorentzian AdS$_{3}$ and then, as needed, Wick rotate to
Euclidean AdS$_{3}$ in order to calculate some physical quantities, but not
to formulate the theory first in Euclidean AdS$_{3}$ and then try to get
Lorentzian theory by Wick rotating back to Lorentzian AdS$_{3}$. In this
paper we will obtain the boundary states for AdS$_{2}$ D-branes directly in
Lorentzian AdS$_{3}$.

The plan of the paper is as follows. In section two we firstly comment on
the background, then we review the analysis in \cite{lomholt} on the
possible gluing conditions that one can obtain from the variation of the
WZNW\ action. In section three we first review the free field formalism of
\cite{bars1} for SL(2,R)\ WZNW\ model and then we translate the gluing
conditions among currents to matching conditions among free fields and later
to the boundary conditions among modes of free fields. At the end of that
section we present our ansatz for the Ishibashi states and show that
boundary conditions imposed on them are satisfied. These states also
preserve one half of the conformal symmetry as expected in a BCFT. In
section four we find the annulus amplitude of closed strings propagating
between two AdS$_{2}$ D-branes. Comparing holomorphic part of this amplitude
to the characters of $\widehat{\mathtt{sl}}$(2,R) Kac Moody algebra we
deduce that the spectrum of closed strings, which couple to AdS$_{2}$
D-branes, are in the principal continuous series of $\widehat{\mathtt{sl}}$%
(2,R) supplemented with zero modes in accordance with the results of \cite
{bars1},\cite{BDM}. In the last sub-section we use the last boundary
condition (\ref{dbc4}) among zero modes to write the physical boundary
states. We conclude with a discussion of possible future research.

\section{Closed and Open Strings in Lorentzian AdS$_{3}$}

\subsection{SL(2,R) WZNW Model and AdS$_{3}$ Space-Time}

Closed string theory on group manifolds are analyzed through WZNW model.
Having affine current algebra (which is nothing but the affinization of the
Lie algebra that is associated to that group manifold) as a symmetry algebra
of the model utilizes the the current algebra technics to find out the
spectrum of the theory.

The WZNW action which describes the closed strings on SL(2,R) manifold is
given by
\begin{equation}
S=\frac{k}{4\pi }\int_{M}d^{2}\sigma \ Tr\left( g^{-1}\partial _{+}g\
g^{-1}\partial _{-}g\right) -\frac{k}{12\pi }\int_{B}d^{3}\varsigma \
Tr\left( g^{-1}dg\ g^{-1}dg\ g^{-1}dg\right) .  \label{WZWaction}
\end{equation}
where $g$ is the SL(2,R) group element and $g^{-1}dg=\left( g^{-1}\partial
_{+}g\right) d\sigma ^{+}+\left( g^{-1}\partial _{-}g\right) d\sigma ^{-}$
is the pull-back on the closed string worldsheet of the left invariant
one-form on the SL(2,R) group manifold.

We parametrize the SL(2,R) group element in the triangular form
\begin{eqnarray}
g\left( z,\bar{z}\right) &=&\left(
\begin{array}{cc}
1 & 0 \\
-\gamma ^{-}\left( z,\bar{z}\right) & 1
\end{array}
\right) \left(
\begin{array}{cc}
e^{\frac{1}{2}\phi \left( z,\bar{z}\right) } & 0 \\
0 & e^{-\frac{1}{2}\phi \left( z,\bar{z}\right) }
\end{array}
\right) \left(
\begin{array}{cc}
1 & -\gamma ^{+}\left( z,\bar{z}\right) \\
0 & 1
\end{array}
\right)  \nonumber \\
&=&\left(
\begin{array}{cc}
e^{\frac{1}{2}\phi } & -\gamma ^{+}e^{\frac{1}{2}\phi } \\
-\gamma ^{-}e^{\frac{1}{2}\phi } & e^{-\frac{1}{2}\phi }+e^{\frac{1}{2}\phi
}\gamma ^{+}\gamma ^{-}
\end{array}
\right) .  \label{figamma}
\end{eqnarray}
where all the fields $\phi \left( z,\bar{z}\right) $, $\gamma ^{-}\left( z,%
\bar{z}\right) $ and $\gamma ^{+}\left( z,\bar{z}\right) $ are real. $\gamma
^{\pm }\left( z,\bar{z}\right) $ are lightcone combinations $\gamma ^{\pm
}=x\pm t$. With this parametrization of the group element the metric on the
group manifold takes the so called Poincar\'{e} form
\begin{equation}
ds^{2}=\frac{k\alpha ^{^{\prime }}}{2}\ Tr\left( g^{-1}dg\ g^{-1}dg\right)
=k\alpha ^{^{\prime }}\left( \frac{1}{u^{2}}du^{2}+u^{2}d\gamma ^{-}d\gamma
^{+}\right)  \label{AdSmetric}
\end{equation}
where $u=e^{\frac{1}{2}\phi \left( z,\bar{z}\right) }$ and $\sqrt{k\alpha
^{^{\prime }}}$ is the radius of curvature. From the the form of the metric
one sees that SL(2,R) group manifold contains one compact time-like
coordinate, $t\in \left[ 0,2\pi \right] $ and two non-compact space-like
coordinates $x\in \left] -\infty ,\infty \right[ $ and $u\in \left[ 0,\infty
\right[ $. To obtain the universal covering space of SL(2,R) group manifold
the time-like coordinate is uncompactified: $t\in \left] -\infty ,\infty
\right[ $. The universal covering space is then the AdS$_{3}$ space-time. In
the present paper we are going to analyze D-branes on this universal
covering space by using the free field formalism of the SL(2,R) WZNW
model. We note that we are not making global identification $x\equiv x+2\pi $%
, i.e. we are not treating the case of vacuum BTZ black hole\footnote{%
Vacuum BTZ black hole with zero mass and zero angular momentum has the same
metric as (\ref{AdSmetric}). However, unlike the Lorentzian AdS$_{3}$
space-time, globally there is the identification $x\equiv x+2\pi $. Closed
string spectrum in such backgrounds is discussed in \cite{troost}.}, but
Lorentzian AdS$_{3}$.

\subsection{Variation of the Action and Gluing Conditions}

The WZNW action (\ref{WZWaction}) describes only the theory of closed
strings on the group manifold. This is rather obvious from the form of the
action: The Wess-Zumino part of the action is an integration over the ball $%
B $ for which the string worldsheet $M$ is the boundary. Since a boundary
cannot have a boundary is a classic result in differential geometry, we
conclude that the WZNW action (\ref{WZWaction}) describes only the closed
strings. Therefore, one cannot work as in the flat Minkowski Space-time. In
flat space-time, the Polyakov action describes both open and closed strings.
Variation of the action results in the equations of motion for both open and
closed strings and in the case of open strings also a surface term, from the
vanishing condition for which one obtains the open string boundary
conditions. Basically these boundary conditions are Neumann boundary
conditions in all directions, which state that momentum does not flow out of
the the string end-points. However, one can play with the kind of boundary
condition in one or all directions. In the T-dual version of flat string
theory, one finds that Neumann boundary condition turns into Dirichlet
boundary condition in the direction that T-duality is performed \cite{TASI}.
Dirichlet boundary condition just means that the string end-points are free
to move in all directions except in the direction for which there is
Dirichlet boundary condition. Therefore, string end-points define a
hyper-surface which is called a Dirichlet-brane (or D-brane). If T-duality
is performed in several directions, the number of directions for which
string end-points obey Dirichlet boundary condition increases and the number
of dimensions that define D-brane decreases. One very important
distinguishing property of D-branes from other hyper-surfaces is that
D-branes have a vector field on their world-volume, which couples to open
string end-points.

If one applies the same kind of logic to the case of string theory on a
group manifold, one has to take into account the above fact that the
hyper-surfaces which are claimed to be D-branes should have a vector field
defined on them. Being careful with this point, Lomholt and Larsen \cite
{lomholt} analyzed the possible boundary conditions that one can write for
possible open strings on SL(2,R) group manifold. In short their analysis is
as follows: They assumed that variation of open string action on SL(2,R)
group manifold is just variation of closed string action (\ref{WZWaction})
plus a term which comes from the coupling of the vector field to the
open-string end-points on possible boundaries. The condition that the
surface term should vanish and also the condition on the field strength of
the vector field (namely the Jacobi condition and that the field strength
should be anti-symmetric) allowed only a few possibilities for the boundary
conditions that one can write for the open strings on SL(2,R) group
manifold. As a result of their analysis, they found that the possible cases
for Dirichlet type boundary conditions are firstly ``regular'' Dirichlet
gluing conditions
\begin{equation}
J\left( z\right) =\tilde{J}\left( \bar{z}\right) \quad at\quad z=\bar{z}
\label{regular}
\end{equation}
and secondly so called ``twisted'' Dirichlet gluing conditions
\begin{equation}
J\left( z\right) =\omega \cdot \tilde{J}\left( \bar{z}\right) \cdot \omega
\quad at\quad z=\bar{z}  \label{twisted}
\end{equation}
where
\[
\omega =\left(
\begin{array}{cc}
0 & 1 \\
1 & 0
\end{array}
\right)
\]
among the currents of $\widehat{\mathtt{sl}}$(2,R) Kac-Moody symmetry.

The regular and twisted gluing conditions above are both written in the open
string channel at $z=\bar{z}$ where $z=e^{i(\tau +\sigma )},\ \bar{z}%
=e^{i(\tau -\sigma )}$. However, we can translate these gluing conditions to
the closed string channel and write them as
\begin{eqnarray}
\mathrm{regular} &:&\quad \left( J(z)+\bar{z}^{2}\,\tilde{J}\left( \bar{z}%
\right) \right) \left| B\right\rangle =0,  \label{reg-closed} \\
\mathrm{twisted} &:&\quad \left( J(z)+\bar{z}^{2}\,\omega \cdot \tilde{J}%
\left( \bar{z}\right) \cdot \omega \right) \left| B\right\rangle =0\quad
at\quad z\cdot \bar{z}=1.  \label{twis-closed}
\end{eqnarray}
The $-\bar{z}^{2}$ factors in front of $\tilde{J}\left( \bar{z}\right) $
comes from the fact that as open string world-sheet with topology of a
cylinder is transformed into a closed string world-sheet with the same
topology one needs to interchange $\sigma $ and $\tau $. Because of the same
reason the boundary in the closed string channel is at $z\cdot \bar{z}=1$.
The state $\left| B\right\rangle $ in (\ref{reg-closed},\ref{twis-closed})
is the boundary state which describes D-branes that constitute the
boundaries for the tree-level amplitude of the propagation of the closed
strings.

The result of the analysis in \cite{lomholt} is in tune with the previous
analysis done in \cite{stanciu},\cite{bachas}. If one expresses the symmetry
currents $J_{L}\left( z\right) =ik\left( \partial g\right) g^{-1}$ and $%
\tilde{J}_{R}\left( \bar{z}\right) =-ikg^{-1}\left( \bar{\partial}g\right) $
in terms of fields $\gamma ^{-}\left( z,\bar{z}\right) $, $\gamma ^{+}\left(
z,\bar{z}\right) $ and $\phi \left( z,\bar{z}\right) $ by using the
representation (\ref{figamma}) of the group element $g$, one finds that
regular gluing condition (\ref{regular}) describes the hyper-surfaces
\begin{equation}
e^{\phi /2}\left( 1+\gamma ^{-}\gamma ^{+}\right) +e^{-\phi /2}=\mathcal{C},
\label{regularD}
\end{equation}
where $\mathcal{C}$ is a arbitrary real constant, as D-branes on AdS$_{3}$.
These D-branes are located on the conjugacy classes of SL(2,R) and depending
on whether $\left| \mathcal{C}\right| $ is bigger, smaller or equal to one,
their world-volumes have the geometry of two dimensional de Sitter space (dS$%
_{2}$), hyperbolic plane (H$_{2}$) or the light cone in three dimensions,
respectively \cite{stanciu},\cite{bachas}. It has been shown that none of
these D-brane world-volumes correspond to a physically acceptable D-brane
motion: for dS$_{2}$ D-branes the Dirac-Born-Infeld action is imaginary and
thus such D-branes are ``unphysical'' \cite{bachas}, for H$_{2}$ D-branes
the world-volume has Euclidean signature and the light-cone breaks up into
three conjugacy classes (apex of the cone, future cone without apex and past
cone without apex) with degenerate induced metric \cite{stanciu},\cite
{bachas}.

Contrary to that, twisted Dirichlet gluing conditions describe physical
D-branes \cite{bachas} which are located on the twisted conjugacy classes of
SL(2,R) \cite{bachas},\cite{lomholt}
\begin{equation}
e^{\phi /2}\left( \gamma ^{-}+\gamma ^{+}\right) =\mathcal{C},
\label{twistedD}
\end{equation}
where $\mathcal{C}$ is a arbitrary real constant, and have the geometry of
AdS$_{2}$ space-time
\begin{equation}
ds^{2}=\frac{1}{u^{2}}du^{2}-u^{2}dt^{2}  \label{ads2}
\end{equation}
where $t=(\gamma ^{+}-\gamma ^{-})/2$.

\section{Free Field Analysis}

In this work our aim is to study the consequences of the twisted Dirichlet
gluing conditions (\ref{twisted}) among Kac-Moody currents and to write the
boundary states for AdS$_{2}$ D-branes so as to determine the spectrum of
closed strings that couple to those AdS$_{2}$ D-branes. We only analyze AdS$%
_{2}$ D-branes, since they are the physical ones, in the present paper. We
are going to write the boundary states as coherent states on SL(2,R) group
manifold. To do that we utilize the free field representation that was used
in \cite{bars1},\cite{BDM} in the study of closed strings on AdS$_{3}$
space-time.

Therefore we now review the free-field representation of the WZNW model on
SL(2,R) manifold as used in \cite{BDM}. With this review we are also going
to present the necessary formulas that will be used to construct the
boundary states for AdS$_{2}$ D-branes on AdS$_{3}$ and fix the notation.

We write the SL(2,R) group element in the following form, separating the
holomorphic and anti-holomorphic parts:
\begin{equation}
g\left( z,\bar{z}\right) =g_{L}\left( z\right) g_{R}\left( \bar{z}\right)
\label{LR}
\end{equation}
\begin{eqnarray}
g_{L}\left( z\right) &=&\left(
\begin{array}{cc}
e^{\frac{1}{2}X_{2}} & -X^{+}e^{\frac{1}{2}X_{2}} \\
-X^{-}e^{\frac{1}{2}X_{2}} & e^{-\frac{1}{2}X_{2}}+e^{\frac{1}{2}%
X_{2}}X^{+}X^{-}
\end{array}
\right)  \label{L} \\
g_{R}\left( \bar{z}\right) &=&\left(
\begin{array}{cc}
e^{\frac{1}{2}\tilde{X}_{2}} & -\tilde{X}^{+}e^{\frac{1}{2}\tilde{X}_{2}} \\
-\tilde{X}^{-}e^{\frac{1}{2}\tilde{X}_{2}} & e^{-\frac{1}{2}\tilde{X}%
_{2}}+e^{\frac{1}{2}\tilde{X}_{2}}\tilde{X}^{+}\tilde{X}^{-}
\end{array}
\right)  \label{R}
\end{eqnarray}
This separation of holomorphic and anti-holomorphic parts is realized when
one derives the equations of motion from the WZNW action. The Kac-Moody
symmetry currents separate into holomorhic and anti-holomorphic parts, which
we call as left and right currents. With the above representation of the
SL(2,R) group element, the left (holomorphic) currents are obtained as%
\footnote{$\tau ^{a}\ (a=1,2,3)$ are the matrix representations of SL(2,R)
generators:$\ $%
\[
\tau ^{0}=\left(
\begin{array}{cc}
0 & 1 \\
-1 & 0
\end{array}
\right) ,\quad \tau ^{1}=\left(
\begin{array}{cc}
0 & 1 \\
1 & 0
\end{array}
\right) ,\quad \tau ^{2}=\left(
\begin{array}{cc}
-1 & 0 \\
0 & 1
\end{array}
\right) .
\]
}:
\begin{equation}
J_{L}\left( z\right) =ik\left( \partial g\right) g^{-1}=ik\left( \partial
g_{L}\right) g_{L}^{-1}=J^{a}\tau ^{a}=\left(
\begin{array}{cc}
-J^{2} & J^{0}+J^{1} \\
-J^{0}+J^{1} & J^{2}
\end{array}
\right)  \label{holocurrent}
\end{equation}
where three generators of holomorphic Kac-Moody algebra are
\begin{eqnarray}
J^{1}(z)+J^{0}(z) &=&P^{+}(z)  \label{wzw1} \\
J^{1}(z)-J^{0}(z) &=&-:X^{-}\,P^{+}X^{-}:-2P_{2}\,X^{-}+i\left( k+2\right)
\partial _{z}X^{-}  \label{wzw2} \\
J_{2}(z) &=&:X^{-}\,P^{+}:+P_{2}(z).  \label{wzw3}
\end{eqnarray}
Here the canonical momenta $P^{+}(z)$ and $P_{2}(z)$ are identified as
\begin{eqnarray}
P^{+}(z) &=&ik\partial _{z}X^{+}e^{X_{2}}  \nonumber \\
P_{2}(z) &=&-\frac{1}{2}ik\partial _{z}X_{2}  \label{canonical}
\end{eqnarray}
Similarly, one calculates the right (anti-holomorphic) currents as
\begin{equation}
\tilde{J}_{R}\left( \bar{z}\right) =-ikg^{-1}\left( \bar{\partial}g\right)
=ikg_{R}^{-1}\left( \bar{\partial}g_{R}\right) =\tilde{J}^{a}\tau
^{a}=\left(
\begin{array}{cc}
-\tilde{J}^{2} & \tilde{J}^{0}+\tilde{J}^{1} \\
-\tilde{J}^{0}+\tilde{J}^{1} & \tilde{J}^{2}
\end{array}
\right)  \label{antiholo}
\end{equation}
where three generators of anti-holomorphic Kac-Moody algebra are
\begin{eqnarray}
\tilde{J}^{1}(\bar{z})-\tilde{J}^{0}(\bar{z}) &=&-\tilde{P}^{-}(\bar{z})
\label{awzw1} \\
\tilde{J}^{1}(\bar{z})+\tilde{J}^{0}(\bar{z}) &=&:\tilde{X}^{+}\,\tilde{P}%
^{-}\,\tilde{X}^{+}:+2\tilde{P}_{2}\,\tilde{X}^{+}-i\left( k+2\right)
\partial _{\bar{z}}\tilde{X}^{+}(\bar{z})  \label{awzw2} \\
\tilde{J}_{2}(\bar{z}) &=&-:\tilde{X}^{+}\,\tilde{P}^{-}:-\tilde{P}_{2}(\bar{%
z})  \label{awzw3}
\end{eqnarray}
The canonical momenta $\tilde{P}^{-}(\bar{z})$ and $\tilde{P}_{2}(\bar{z})$
are identified as
\begin{eqnarray}
\tilde{P}^{-}(\bar{z}) &=&ik\partial _{\bar{z}}\tilde{X}^{-}e^{\tilde{X}_{2}}
\nonumber \\
\tilde{P}_{2}(\bar{z}) &=&-\frac{1}{2}ik\partial _{\bar{z}}\tilde{X}_{2}
\label{acanonical}
\end{eqnarray}
In (\ref{wzw1}-\ref{wzw3}) the holomorphic fields $X^{-}\left( z\right) $, $%
P^{+}\left( z\right) $ and $P_{2}\left( z\right) $ are free fields and have
the following mode expansions:
\begin{eqnarray}
X^{-}\left( z\right) &=&q^{-}-ip^{-}\ln z+\sum_{n\neq 0}\frac{1}{n}\alpha
_{n}^{-}z^{-n},\qquad \left( \alpha _{n}^{-}\right) ^{\dagger }=\alpha
_{-n}^{-}\quad  \label{x-} \\
P^{+}\left( z\right) &=&\frac{p^{+}}{z}+\sum_{n\neq 0}\alpha
_{n}^{+}z^{-n-1},\qquad \qquad \qquad \left( \alpha _{n}^{+}\right)
^{\dagger }=\alpha _{-n}^{+}  \label{p+} \\
P_{2}\left( z\right) &=&\frac{p_{2}}{z}+\sum_{n\neq 0}s_{n}z^{-n-1},\qquad
\qquad \qquad \quad \left( s_{n}\right) ^{\dagger }=s_{-n}\ .  \label{p2}
\end{eqnarray}
The fields ($X^{-},P^{+})$ are conjugate to each other and similarly $\left(
X_{2},P_{2}\right) $. The modes of the free fields obey the following
commutation rules:
\begin{equation}
\begin{array}{l}
\lbrack q^{-},p^{+}]=i\,, \\
\left[ \alpha _{n}^{-}\,,\alpha _{m}^{+}\right] =n\,\delta _{n+m,0}\,\,\,,
\\
\left[ s_{n}\,,s_{m}\right] =\left( \frac{k}{2}-1\right) \,n\,\delta
_{n+m,0}\,\,\,.
\end{array}
\label{commutators}
\end{equation}
Likewise anti-holomorphic free-fields $\tilde{X}^{+}\left( \bar{z}\right) $,
$\tilde{P}^{+}\left( \bar{z}\right) $ and $\tilde{P}_{2}\left( \bar{z}%
\right) $ have mode expansions:
\begin{eqnarray}
\tilde{X}^{+}\left( \bar{z}\right) &=&\tilde{q}^{+}-i\tilde{p}^{+}\ln \bar{z}%
+\sum_{n\neq 0}\frac{1}{n}\tilde{\alpha}_{n}^{+}\bar{z}^{-n},\qquad \left(
\tilde{\alpha}_{n}^{+}\right) ^{\dagger }=\tilde{\alpha}_{-n}^{+}\quad
\label{x+} \\
\tilde{P}^{-}\left( \bar{z}\right) &=&\frac{\tilde{p}^{-}}{z}+\sum_{n\neq 0}%
\tilde{\alpha}_{n}^{-}\bar{z}^{-n-1},\qquad \qquad \qquad \left( \tilde{%
\alpha}_{n}^{-}\right) ^{\dagger }=\tilde{\alpha}_{-n}^{-}  \label{p-} \\
\tilde{P}_{2}\left( \bar{z}\right) &=&\frac{\tilde{p}_{2}}{z}+\sum_{n\neq 0}%
\tilde{s}_{n}\bar{z}^{-n-1},\qquad \qquad \qquad \ \ \left( \tilde{s}%
_{n}\right) ^{\dagger }=\tilde{s}_{-n}\ .  \label{ap2}
\end{eqnarray}
Here the fields ($\tilde{X}^{+},\tilde{P}^{-})$ are conjugate to each other
and similarly $\left( \tilde{X}_{2},\tilde{P}_{2}\right) $. The commutation
relations among modes are
\begin{equation}
\begin{array}{l}
\lbrack \tilde{q}^{+},\tilde{p}^{-}]=i\,, \\
\left[ \tilde{\alpha}_{n}^{+}\,,\tilde{\alpha}_{m}^{-}\right] =n\,\delta
_{n+m,0}\,\,\,, \\
\left[ \tilde{s}_{n}\,,\tilde{s}_{m}\right] =\left( \frac{k}{2}-1\right)
\,n\,\delta _{n+m,0}\,\,.\,
\end{array}
\label{acommutators}
\end{equation}

The holomorphic Sugawara energy-momentum tensor takes the form (after
careful ordering of operators, including zero modes, to insure hermiticity)
\begin{equation}
T\left( z\right) =:P^{+}i\partial X^{-}:+\frac{1}{k-2}\left( :P_{2}^{2}:-%
\frac{i}{z}\partial \left( zP_{2}\right) +\frac{1}{4z^{2}}\right)
\label{T++}
\end{equation}
From this expression for the energy-momentum tensor, by using the mode
expansion of holomorphic free fields, we obtain the Virasoro generators in
the holomorphic sector as
\begin{equation}
L_{n}=\sum_{m=-\infty }^{\infty }:\alpha _{n-m}^{+}\alpha _{m}^{-}:+\frac{1}{%
k-2}\left[ \sum_{m=-\infty }^{\infty }:s_{n-m}s_{m}:+ins_{n}+\frac{\delta
_{n,0}}{4}\right]  \label{L_n}
\end{equation}
where $\alpha _{0}^{+}=p^{+}$, $\alpha _{0}^{-}=p^{-}$ and $s_{0}=p_{2}$.

The anti-holomorphic energy-momentum tensor and the Virasoro generators have
the same form as the above expressions.

\subsection{Matching Conditions}

After this review of the free field realization of the WZNW model on SL(2,R)
group manifold now we are ready to analyze the consequences of the twisted
Dirichlet gluing conditions (\ref{twisted}) among Kac-Moody currents. We
will use the expressions (\ref{holocurrent}) and (\ref{antiholo}) to
translate the twisted gluing conditions among the symmetry currents to
boundary conditions among the free fields. The boundary conditions among the
free fields are called as ``matching conditions'' \cite{ishikawa}.
Subsequently from the matching conditions, by using the mode expansions of
the free fields, the ``boundary conditions'' among the modes will be derived.

Using (\ref{twisted}), (\ref{holocurrent}) and (\ref{antiholo}) the twisted
gluing conditions in open string channel at $z=\bar{z}$ become
\begin{eqnarray}
-J^{2} &=&\tilde{J}^{2} \\
J^{1}-J^{0} &=&\tilde{J}^{1}+\tilde{J}^{0} \\
J^{1}+J^{0} &=&\tilde{J}^{1}-\tilde{J}^{0}
\end{eqnarray}
and in the closed string channel at $z\cdot \bar{z}=1$ they become
\begin{eqnarray}
\left[ J^{2}-\bar{z}^{2}\ \tilde{J}^{2}\right] \left| B\right\rangle &=&0 \\
\left[ \left( J^{1}-J^{0}\right) +\bar{z}^{2}\ \left( \tilde{J}^{1}+\tilde{J}%
^{0}\right) \right] \left| B\right\rangle &=&0 \\
\left[ \left( J^{1}+J^{0}\right) +\bar{z}^{2}\ \left( \tilde{J}^{1}-\tilde{J}%
^{0}\right) \right] \left| B\right\rangle &=&0.
\end{eqnarray}

Substituting in the above gluing conditions the free field representations (%
\ref{wzw1}-\ref{wzw3}, \ref{awzw1}-\ref{awzw3}) of symmetry currents we
obtain the matching conditions among free fields in the open string channel
as
\begin{equation}
P^{+}\left( z\right) =-\tilde{P}^{-}\left( \bar{z}\right) ,\quad P_{2}\left(
z\right) =\tilde{P}_{2}\left( \bar{z}\right) ,\quad X^{-}\left( z\right) =-%
\tilde{X}^{+}\left( \bar{z}\right) \quad at\quad z=\bar{z}
\end{equation}
and in the closed string channel they are
\begin{eqnarray}
\left[ P^{+}\left( z\right) -\bar{z}^{2}\ \tilde{P}^{-}\left( \bar{z}\right)
\right] \left| B\right\rangle &=&0, \\
\quad \left[ P_{2}\left( z\right) +\bar{z}^{2}\ \tilde{P}_{2}\left( \bar{z}%
\right) \right] \left| B\right\rangle &=&0, \\
\quad \left[ X^{-}\left( z\right) +\tilde{X}^{+}\left( \bar{z}\right)
\right] \left| B\right\rangle &=&0\quad at\quad z\cdot \bar{z}=1.
\end{eqnarray}

Next we use the mode expansions of the free fields, (\ref{x-},\ref{p+},\ref
{p2}) and (\ref{x+},\ref{p-},\ref{ap2}), to find the boundary conditions
among the modes of free fields in the closed string channel. In terms of
modes the twisted Dirichlet boundary condition reads in the closed string
channel as
\begin{eqnarray}
\underline{\mathrm{Zero\ Modes}}\quad \quad &&\underline{\mathrm{Oscillator\
Modes}}  \nonumber \\
\left( p^{+}-\tilde{p}^{-}\right) \left| B\right\rangle =0\quad \quad
&&\left( \alpha _{n}^{+}-\tilde{\alpha}_{-n}^{-}\right) \left|
B\right\rangle =0  \label{dbc1} \\
\left( p_{2}+\tilde{p}_{2}\right) \left| B\right\rangle =0\quad \quad
&&\left( s_{n}+\tilde{s}_{-n}\right) \left| B\right\rangle =0  \label{dbc2}
\\
\left( p^{-}-\tilde{p}^{+}\right) \left| B\right\rangle =0\quad \quad
&&\left( \alpha _{n}^{-}-\tilde{\alpha}_{-n}^{+}\right) \left|
B\right\rangle =0  \label{dbc3} \\
\left( q^{-}+\tilde{q}^{+}\right) \left| B\right\rangle =0\quad \quad &&
\label{dbc4}
\end{eqnarray}

\subsection{Coherent Boundary States}

The free field analysis of the previous subsections allows us to write the
spectrum of closed strings on Lorentzian AdS$_{3}$ as a Fock space built on
a base state which is a direct product of base states for holomorphic and
anti-holomorphic sectors \cite{bars1}:
\begin{equation}
\left| base\right\rangle =\left| p^{+},p^{-},p_{2}\right\rangle \otimes
\left| \ \tilde{p}^{+},\tilde{p}^{-},\tilde{p}_{2}\right\rangle .
\label{base1}
\end{equation}
Then the physical states are obtained by applying creation oscillators $%
\alpha _{-n}^{+},\ \alpha _{-n}^{-},\ s_{-n},\ \left( n>0\right) $ and their
anti-holomorhic counterparts on this base state.

Our aim is to find conformally invariant boundary states built on the above
base state. In the case of bosonic string theory on flat Minkowski
space-time such states were found in the form of coherent states \cite
{ishibashi1}. Hence, we have the following ansatz for the coherent boundary
state:
\begin{eqnarray}
\left| B_{p^{+},p^{-},p_{2}}\right\rangle &=&\exp \left( \sum_{n>0}\frac{1}{n%
}\alpha _{-n}^{-}\tilde{\alpha}_{-n}^{-}+\sum_{n>0}\frac{1}{n}\alpha
_{-n}^{+}\tilde{\alpha}_{-n}^{+}-\sum_{n>0}\frac{1}{n}s_{-n}^{^{\prime }}%
\tilde{s}_{-n}^{^{\prime }}\right) \left| p^{+},p^{-},p_{2}\right\rangle
\label{Ishibashi1} \\
\left\langle B_{p^{+},p^{-},p_{2}}\right| &=&\left\langle
p^{+},p^{-},p_{2}\right| \exp \left( \sum_{n>0}\frac{1}{n}\alpha _{n}^{-}%
\tilde{\alpha}_{n}^{-}+\sum_{n>0}\frac{1}{n}\alpha _{n}^{+}\tilde{\alpha}%
_{n}^{+}-\sum_{n>0}\frac{1}{n}s_{n}^{^{\prime }}\tilde{s}_{n}^{^{\prime
}}\right)  \label{Ishibashi2}
\end{eqnarray}
where $s_{n}^{^{\prime }}=\frac{s_{n}}{\sqrt{\frac{k}{2}-1}}$ and $\tilde{s}%
_{n}^{^{\prime }}=\frac{\tilde{s}_{n}}{\sqrt{\frac{k}{2}-1}}$. In this
ansatz for coherent boundary state we have taken the base state as
\begin{equation}
\left| base\right\rangle =\left| p^{+},p^{-},p_{2}\right\rangle
\label{base2}
\end{equation}
already using the twisted Dirichlet boundary conditions for zero modes (\ref
{dbc1}-\ref{dbc3}). We have to check whether this coherent boundary states
obey also the twisted Dirichlet boundary conditions for oscillator modes (%
\ref{dbc1}-\ref{dbc3}). Indeed, they satisfy:
\begin{eqnarray}
(\alpha _{n}^{+}-\tilde{\alpha}_{-n}^{-})\left|
B_{p^{+},p^{-},p_{2}}\right\rangle &=&0=\left\langle
B_{p^{+},p^{-},p_{2}}\right| (\alpha _{n}^{+}-\tilde{\alpha}_{-n}^{-}), \\
(\alpha _{n}^{-}-\tilde{\alpha}_{-n}^{+})\left|
B_{p^{+},p^{-},p_{2}}\right\rangle &=&0=\left\langle
B_{p^{+},p^{-},p_{2}}\right| (\alpha _{n}^{-}-\tilde{\alpha}_{-n}^{+}), \\
(s_{n}+\tilde{s}_{-n})\left| B_{p^{+},p^{-},p_{2}}\right\rangle
&=&0=\left\langle B_{p^{+},p^{-},p_{2}}\right| (s_{n}+\tilde{s}_{-n}).
\end{eqnarray}
In BCFT such boundary states, that obey the necessary boundary conditions,
are called as Ishibashi states \cite{ishibashi2}.

To have a consistent BCFT the conformal symmetry of the theory should be
preserved. This means that, in the open string channel, the holomorphic and
anti-holomorphic energy-momentum tensors should match at the boundary
\begin{equation}
\left[ T\left( z\right) -\tilde{T}\left( \bar{z}\right) \right] =0\quad
at\quad z=\bar{z},  \label{TTbar}
\end{equation}
since the energy-momentum tensor is the generator of the conformal
transformations. In the closed string channel this condition becomes a
relation between holomorphic and anti-holomorphic Virasoro generators
\begin{equation}
(L_{n}-\tilde{L}_{-n})\left| B_{_{p^{+},p^{-},p_{2}}}\right\rangle
=0=\left\langle B_{p^{+},p^{-},p_{2}}\right| (L_{n}-\tilde{L}_{-n}).
\label{confsym}
\end{equation}
Our coherent boundary states (\ref{Ishibashi1}-\ref{Ishibashi2}) also obey
these conditions.

The last twisted Dirichlet boundary condition (\ref{dbc4}) will be used in
the next section in order to determine the form of the ``physical'' boundary
states and the location of AdS$_{2}$ D-branes. That is this condition will
be imposed on the physical boundary states not on the Ishibashi states.

\section{Physical Boundary States}

\subsection{Annulus Amplitude and Closed String Spectrum}

Any combination of Ishibashi states
(\ref{Ishibashi1},\ref{Ishibashi2}) obey the twisted Dirichlet
boundary conditions (\ref{dbc1}-\ref{dbc3}). However, not all the
combinations can be taken as genuine boundary state. Apart from
the gluing conditions (\ref{dbc1}-\ref{dbc3}) and the condition to
preserve the conformal symmetry, there is an extra condition which
chooses the physical boundary states among the infinitely many
combinations of the Ishibashi states. The extra condition comes
from the fact that the annulus amplitude in the closed string
channel, which is just the tree-level amplitude of propagation of
a closed string from one boundary $B$ to the other
boundary $B^{\prime }$\footnote{%
Since in the representation of Kac-Moody currents the diagonal generator is $%
J^{+}(z)$, we are using its zero mode $J_{0}^{+}$ in the expression for the
annulus amplitude.}
\begin{equation}
\left\langle B_{p^{+\prime },p^{-\prime },p_{2}^{\prime }}^{^{\prime
}}\right| \ e^{2\pi i\tau (L_{0}-\frac{c}{24})}e^{-2\pi i\bar{\tau}(\tilde{L}%
_{0}-\frac{c}{24})}\ e^{2\pi i\theta J_{0}^{+}}e^{-2\pi i\bar{\theta}\tilde{J}%
_{0}^{+}}\ \left| B_{p^{+},p^{-},p_{2}}\right\rangle  \label{annulus}
\end{equation}
should be equivalent to the one-loop amplitude in the open string channel
with boundary conditions $B$ and $B^{\prime }$. This condition is first
discussed by Cardy in the case of rational conformal field theories (RCFT)
in \cite{cardy} and therefore it is known as Cardy condition. However, since
SL(2,R) WZNW model contains infinite number of primary fields and,
therefore, is not a RCFT, one cannot use the Cardy condition directly.
Either one has to utilize an analog of the Cardy condition for non-RCFT as
in \cite{teschner-0} and \cite{teschner} or try another way in order to
determine the physical boundary states. In this paper we do not use a
(modified) Cardy condition, instead we determine the physical boundary
states using the condition (\ref{dbc4}) between the zero modes of $X^{-}$
and $\tilde{X}^{+}$.\footnote{%
Similar analysis was done in the case of SU(2) WZW model in \cite{ishikawa}.}

However, before determining the physical boundary states with the above
mentioned method we would like to derive the annulus amplitude in the free
field formalism and use it to decide the spectrum of the closed strings that
couple to the AdS$_{2}$ D-branes on Lorentzian AdS$_{3}$. The holomorphic
part of the annulus amplitude
\begin{equation}
\hat{\chi}_{_{p^{+},p^{-},p_{2}}}\left( \tau ,\theta \right)
=\left\langle
B_{p^{+\prime },p^{-\prime },p_{2}^{\prime }}^{^{\prime }}\right| \ q^{L_{0}-%
\frac{c}{24}}\ w^{J_{0}^{+}}\ \left| B_{p^{+},p^{-},p_{2}}\right\rangle
\label{ffchar}
\end{equation}
where $q=\exp \left( 2\pi i\tau \right) ,\ w=\exp \left( 2\pi
i\theta \right) $,
can be calculated straightforwardly by separating the Virasoro generator $%
L_{0}$ into zero-mode and oscillator parts
\begin{equation}
L_{0}=p^{+}p^{-}+\frac{1}{k-2}\left( p_{2}^{2}+\frac{1}{4}\right)
+\sum_{m>0}\left( \alpha _{-m}^{+}\alpha _{m}^{-}+\alpha _{-m}^{-}\alpha
_{m}^{+}+s_{n-m}^{\prime }s_{m}^{\prime }\right)  \label{L0}
\end{equation}
where $s_{n}^{^{\prime }}=\frac{s_{n}}{\sqrt{\frac{k}{2}-1}}$ and $\tilde{s}%
_{n}^{^{\prime }}=\frac{\tilde{s}_{n}}{\sqrt{\frac{k}{2}-1}}$. Then one uses
the coherent state methods to obtain the contribution of oscillators to the
free field character (\ref{ffchar}) as
\begin{equation}
\prod_{n>0}\left( \frac{1}{1-q^{n}}\right) ^{3}=\frac{q^{1/8}}{\eta \left(
\tau \right) ^{3}}\ .
\end{equation}
The contribution from zero-modes is
\begin{equation}
\left\langle p^{+\prime },p^{-\prime },p_{2}^{\prime }\right| \ q^{\left(
p^{+}p^{-}+\frac{1}{k-2}\left( p_{2}^{2}+\frac{1}{4}\right) -\frac{k}{%
8\left( k-2\right) }\right) }\ w^{p^{+}}\ \left|
p^{+},p^{-},p_{2}\right\rangle =q^{\left( p^{+}p^{-}+\frac{p_{2}^{2}}{k-2}-%
\frac{1}{8}\right) }\ w^{p^{+}}.
\end{equation}
Combining them one finds the full free-field character:
\begin{equation}
\hat{\chi}_{_{p^{+},p^{-},p_{2}}}\left( \tau ,\theta \right)
=\frac{1}{\eta \left( \tau \right) ^{3}}\ q^{\left(
p^{+}p^{-}+\frac{p_{2}^{2}}{k-2}\right) }\ w^{p^{+}}.
\label{fullffchar}
\end{equation}

In order to find the spectrum of the closed string states which couple to AdS%
$_{2}$ D-branes we compare the holomorhic part of the annulus amplitude (or
free field character) with the characters of the irreducible representations
of $\widehat{\mathtt{sl}}$(2,R). This is because if we had written down the
Ishibashi states in a specific representation of $\widehat{\mathtt{sl}}$%
(2,R) instead of free field representation, the holomorphic part
of the annulus amplitude in the closed string channel would be
just the character of $\widehat{\mathtt{sl}}$(2,R) for that
specific representation.\footnote{%
See \cite{rajaraman} for a form of such Ishibashi states (equ.
(20) in that paper). However, even though in \cite{rajaraman} it
is claimed that only the states in the principal continuous series
representation of $\widehat{\mathtt{sl}}$(2,R) are treated, the
annulus amplitude (equ. (26) in \cite{rajaraman}) is in the form
of discrete series character of $\widehat{\mathtt{sl}}$(2,R),
which is in contradiction with the claim made.} The free field
character (\ref{fullffchar}) is most closely related to the
character of
the principal continuous series representation, $\hat{C}_{j}^{\alpha }$, of $%
\widehat{\mathtt{sl}}$(2,R) \cite{roberts}:
\begin{equation}
\hat{\chi}_{_{j}}\left( \tau ,\theta \right) =\frac{1}{\eta \left(
\tau \right) ^{3}}\ q^{\left( \frac{\rho ^{2}}{k-2}\right)
}\sum_{m=-\infty }^{\infty }w^{m+\alpha }
\end{equation}
where $j=-\frac{1}{2}+i\rho $ and $\alpha \in \left[ 0,1\right) $.
Therefore, we expect that the states of closed strings, which couple to the
AdS$_{2}$ D-branes in Lorentzian AdS$_{3}$, belong to the principal
continuous series of $\widehat{\mathtt{sl}}$(2,R). Since the free field
character (\ref{fullffchar}) differs from the character of principal
continuous series of $\widehat{\mathtt{sl}}$(2,R), we conjecture the
``consistent'' boundary states\footnote{%
Here by consistent boundary states, unlike its usage in \cite{kawai}, we do
not mean Cardy states. In order to find Cardy states we still have to use
boundary condition among zero modes (\ref{dbc4}) and we will call the final
result as the ``physical'' boundary state instead of the Cardy state,
because we are not using Cardy(--like) condition in our analysis.} as sum of
``coherent'' boundary states as
\begin{equation}
\left| \mathcal{B}\right\rangle =\sum_{m\in Z}\ \sum_{r\in Z_{+}}\left|
B_{m+\alpha ,-\frac{r}{m+\alpha },p_{2}}\right\rangle  \label{consistent}
\end{equation}
where we set $p^{+}=m+\alpha $, $m\in Z$ and $p^{+}p^{-}=-r$, $r\in Z_{+}$.
The second relation originates from the monodromy considerations \cite{bars1}
and is also reviewed below. With this form of the consistent boundary states
the holomorphic part of the annulus amplitude become
\begin{equation}
\hat{\chi}\left( \tau ,\theta \right) =\left\langle
\mathcal{B}^{^{\prime
}}\right| \ q^{L_{0}-\frac{c}{24}}\ w^{J_{0}^{+}}\ \left| \mathcal{B}%
\right\rangle =\frac{1}{\eta \left( \tau \right) ^{3}}\ q^{\left( \frac{%
p_{2}^{2}}{k-2}\right) }\sum_{r\in Z_{+}}\ q^{-r}\sum_{m\in Z}\ w^{m+\alpha }
\label{holo-amp}
\end{equation}
If one sets $p^{-}=0$ and $p_{2}=\rho $ one obtains the character for the
principal continuous series\footnote{%
Since these expressions are divergent, we are just comparing, formally, the
form of the expressions.}. However, as explained below one should not set $%
p^{-}=0$. Hence, we conclude that the closed string states which couple to
the AdS$_{2}$ D-branes in Lorentzian AdS$_{3}$ are in the principal
continuous series of $\widehat{\mathtt{sl}}$(2,R) supplemented with the zero
mode $p^{-}$.

The form of the holomorphic part of the annulus amplitude gives support to
the conclusions in papers \cite{bars1},\cite{BDM}, namely that the closed
string spectrum on SL(2,R) group manifold contains only the states from the
principal continuous series of $\widehat{\mathtt{sl}}$(2,R) Kac-Moody
algebra supplemented with the zero mode $p^{-}$. In the anti-holomorphic
sector there is also another extra degree of freedom $\tilde{p}^{+}$, which
is equivalent to $p^{-}$ in the open string channel. The effect of this
extra degree of freedom in the holomorphic sector is that it contributes a
new term to the mass-shell condition \cite{bars1},\cite{BDM}. Without $p^{-}$
the eigenvalue of the Virasoro generator $L_{0}$ (Hamiltonian) on a physical
state at level $N$ is
\begin{equation}
L_{0}\left| phys\right\rangle =\left( -\frac{j\left( j+1\right) }{k-2}%
+N\right) \left| phys\right\rangle \,.  \label{L0-discrete}
\end{equation}
Mass-shell condition requires that $L_{0}=a\leq 1$. In order to obey this
condition, since $N$ is positive, $j\left( j+1\right) $ should be positive.
This requires that only the highest and the lowest weight discrete series
representations of $\widehat{\mathtt{sl}}$(2,R), for which $j\left(
j+1\right) >-\frac{1}{4}$, should be taken into account. Other than these
representations of $\widehat{\mathtt{sl}}$(2,R) only the lowest level states
from the principal continuous series are allowed. However, in the presence
of $p^{-}$ the eigenvalue of $L_{0}$ is modified as
\begin{equation}
L_{0}\left| phys\right\rangle =\left( p^{+}p^{-}-\frac{j\left( j+1\right) }{%
k-2}+N\right) \left| phys\right\rangle .  \label{L0-principal}
\end{equation}
Considerations of monodromy (i.e. invariance of closed string states under
transformation $\sigma \rightarrow \sigma +2\pi $ on closed string
world-sheet) have shown in \cite{bars1} that $p^{+}p^{-}$ should be equal to
a negative integer
\begin{equation}
p^{+}p^{-}=-r,\quad r\in Z_{+}
\end{equation}
Since $p^{+}p^{-}$ is negative and $N$ is positive, now on shell condition $%
L_{0}=a\leq 1$ can be satisfied with $j\left( j+1\right) $ being negative.
This means that the physical states of closed strings on SL(2,R) group
manifold belong to the principal continuous series of $\widehat{\mathtt{sl}}$%
(2,R) Kac-Moody algebra. Existence or non-existence of $p^{-}$, therefore,
changes completely the spectrum of physical states. Which of these
mass-shell conditions, and consequently which spectrum is the physical
spectrum can be deduced by going to the flat limit in the SL(2,R) WZNW\
model. The flat limit of the model is obtained by sending the radius of
curvature of AdS$_{3}$ to infinity. From the metric of AdS$_{3}$ (\ref
{AdSmetric}) the radius of curvature is read as $R=\sqrt{k\alpha ^{^{\prime
}}}$. Therefore $k\rightarrow \infty $ limit in SL(2,R) WZNW model is the
flat limit that we are seeking for.

It has been shown in \cite{BDM} that the vertex operator for the lowest
level closed string states in the momentum basis is
\begin{equation}
V_{p^{+},p^{-}}^{r,p_{2}}\left( z\right) =Ce^{iX^{-}p^{+}}\,\,e^{-\frac{1}{2}%
X_{2}}\,J_{2\sigma }\left( 2\sqrt{-p^{+}p^{-}}e^{-\frac{1}{2}%
X_{2}(z)}\right) \,e^{iX^{+}p^{-}}  \label{braket}
\end{equation}
where $r\in Z_{+}$, $\sigma =\sqrt{kr-p_{2}^{2}}$ and $C$ is an overall
factor. In the flat limit, $k\rightarrow \infty $, this vertex operator
becomes the correct vertex operator for the lowest level closed string
states in Minkowski space-time
\begin{equation}
\left( V_{p^{+}p^{-}}^{r,p_2}(z)\right) _{k\rightarrow \infty }=e^{i\breve{X}%
^{-}\breve{p}^{+}}e^{i\breve{X}^{2}\breve{p}^{2}}e^{i\breve{X}^{+}\breve{p}%
^{-}}\delta \left( \breve{p}^{+}\breve{p}^{-}+\breve{p}_{2}^{2}+m^{2}\right)
\label{flat-vertex}
\end{equation}
where $m=\frac{\sigma }{\sqrt{k}}$ is mass, $\breve{p}^{+}=\frac{1}{\sqrt{k}}%
p^{+}$, $\breve{p}^{-}=\frac{1}{\sqrt{k}}p^{-}$, $\breve{p}_{2}=\frac{1}{%
\sqrt{k}}p_{2}$ are flat space-time momenta and $\breve{X}^{-}=\sqrt{k+2}%
\,X^{-}$, $\breve{X}^{+}=\sqrt{k+2}\,X^{+}$, $\breve{X}_{2}=\sqrt{k+2}%
\,X_{2} $ are holomorphic part of the fields on closed string world-sheet in
flat Minkowski space-time. In the expression for flat vertex operator (\ref
{flat-vertex}) we also see the mass-shell condition in flat space-time:
\begin{equation}
\breve{p}^{+}\breve{p}^{-}+\breve{p}_{2}^{2}+m^{2}=0.
\label{flat-mass-shell}
\end{equation}
It is absolutely necessary to have flat momentum $\breve{p}^{-}$ in this
expression in order to have physically acceptable motion of any relativistic
particle or string in the flat Minkowski space-time. Therefore it is
absolutely necessary to have its corresponding quantity $p^{-}$ in the
formulation of string theory in curved AdS$_{3}$ space-time.

In this paper we have reached this same conclusion from the viewpoint of AdS$%
_{2}$ D-branes: the correct spectrum of closed strings in Lorentzian AdS$%
_{3} $ space-time is the principal continuous series of $\widehat{\mathtt{sl}%
}$(2,R) Kac-Moody algebra supplemented with zero modes.

\subsection{Location of AdS$_{2}$ D-branes}

In section 3.2 we have shown that the coherent boundary states (\ref
{Ishibashi1},\ref{Ishibashi2}) obey the twisted Dirichlet type boundary
conditions (\ref{dbc1}-\ref{dbc3}) and therefore give a representation of
Ishibashi states. These states also preserve one half of the conformal
symmetry (\ref{confsym}). The only condition that is left to be satisfied is
the condition among the zero modes
\begin{equation}
\left( q^{-}+\tilde{q}^{+}\right) \left| B\right\rangle =0
\end{equation}
which contains information about the location of AdS$_{2}$ D-branes in AdS$%
_{3}$ space-time. However, this condition has been derived by an algebraic
analysis and thus without using any extra data on the group manifold the
above condition only applies to AdS$_{2}$ D-brane at the unit element of
SL(2,R).

It has been shown in \cite{stanciu-0},\cite{bachas} that the twisted
Dirichlet boundary condition (\ref{dbc4}) --when translated from the unit
element of the group to an arbitrary point on the group manifold-- requires
the corresponding D-branes to be located on twisted conjugacy classes. Given
the particular representation of an arbitrary group element of SL(2,R) (\ref
{figamma}), the twisted conjugacy classes of SL(2,R) are given by the
condition \cite{stanciu},\cite{bachas}
\begin{equation}
e^{\phi /2}\left( \gamma ^{-}+\gamma ^{+}\right) =\mathcal{C},
\label{twistedD2}
\end{equation}
where $-\infty <\mathcal{C}<\infty $ is a real constant. Let us analyze this
condition in the free field representation of the SL(2,R) group element $g$ (%
\ref{LR}-\ref{R}). Equating expressions (\ref{figamma}) and (\ref{LR}) for $%
g $ we find \cite{BDM}
\begin{eqnarray}
\phi \left( z,\bar{z}\right) &=&X_{2}\left( z\right) +\tilde{X}_{2}\left(
\bar{z}\right) +2\ln \left( 1+X^{+}\left( z\right) \tilde{X}^{-}\left( \bar{z%
}\right) \right)  \label{ads11} \\
\gamma ^{-}\left( z,\bar{z}\right) &=&X^{-}\left( z\right) +\frac{%
e^{-X_{2}\left( z\right) }\tilde{X}^{-}\left( \bar{z}\right) }{1+X^{+}\left(
z\right) \tilde{X}^{-}\left( \bar{z}\right) }  \label{ads12} \\
\gamma ^{+}\left( z,\bar{z}\right) &=&\tilde{X}^{+}\left( \bar{z}\right) +%
\frac{e^{-\tilde{X}_{2}\left( \bar{z}\right) }X^{+}\left( z\right) }{%
1+X^{+}\left( z\right) \tilde{X}^{-}\left( \bar{z}\right) }\ .  \label{ads13}
\end{eqnarray}
Substituting the expressions for $\gamma ^{-}\left( z,\bar{z}\right) $, $%
\gamma ^{+}\left( z,\bar{z}\right) $ and $e^{\phi \left( z,\bar{z}\right)
/2} $ into (\ref{twistedD2}) and then using the matching condition $%
X^{-}(z)=-\tilde{X}^{+}(\bar{z})$ at $z=\bar{z}$ in the open string channel
we obtain
\begin{equation}
e^{-X_{2}\left( z\right) }\tilde{X}^{-}\left( \bar{z}\right) +e^{-\tilde{X}%
_{2}\left( \bar{z}\right) }X^{+}\left( z\right) -\mathcal{C\,}%
e^{-X_{2}\left( z\right) -\tilde{X}_{2}\left( \bar{z}\right) }=0.
\label{twistedD3}
\end{equation}
In the free field approach, that we utilize, $X^{+}(z)$ and $\tilde{X}^{-}(%
\bar{z})$ are not free fields, but are composites of free fields \cite{bars1}%
,\cite{BDM}:
\begin{eqnarray}
X^{+}\left( z\right) &=&q^{+}-\frac{i}{k}\int^{z}dz^{\prime }P^{+}(z^{\prime
})e^{-X_{2}\left( z^{\prime }\right) },\quad \\
\tilde{X}^{-}\left( \bar{z}\right) &=&\tilde{q}^{-}-\frac{i}{k}\int^{\bar{z}%
}d\bar{z}^{\prime }\tilde{P}^{-}(\bar{z}^{\prime })\,e^{-\tilde{X}_{2}\left(
\bar{z}^{\prime }\right) }.
\end{eqnarray}
Substituting these expressions into (\ref{twistedD3}) and using matching
conditions
\begin{eqnarray*}
P^{+}\left( z\right) &=&-\tilde{P}^{-}\left( \bar{z}\right) , \\
P_{2}\left( z\right) &=&\tilde{P}_{2}\left( \bar{z}\right) \qquad
\Rightarrow \quad X_{2}(z)=\tilde{X}_{2}(\bar{z})+\left( q_{2}-\tilde{q}%
_{2}\right)
\end{eqnarray*}
we obtain
\begin{equation}
q_{2}-\tilde{q}_{2}=\Theta .  \label{location}
\end{equation}
where $\Theta $ is a function of $q^{+},\ \tilde{q}^{-}$ and $\mathcal{C}$
and can be obtained by solving (\ref{twistedD3}). In order to write the
physical boundary states we have to construct appropriate combinations of
consistent boundary states (\ref{consistent}) and form a $\delta $--function
realizing the boundary condition (\ref{location}) among the zero modes. This
$\delta $--function can be represented as
\begin{equation}
\delta (q_{2}-\tilde{q}_{2}-\Theta )=\int\limits_{-\infty }^{\infty }d\rho
\,e^{i\rho (q_{2}-\tilde{q}_{2}-\Theta )}=\int\limits_{-\infty }^{\infty
}d\rho \,e^{-i\rho \Theta }\,e^{i\rho (q_{2}-\tilde{q}_{2})}.  \label{delta}
\end{equation}
and it describes the wave function of the zero modes. Then, the physical
boundary state for AdS$_{2}$ D-branes can be written as
\begin{equation}
\left| \mathcal{B}_{phys}\right\rangle =\int\limits_{-\infty }^{\infty
}d\rho \sum_{m\in Z}\ \sum_{r\in Z_{+}}\,e^{-i\rho \Theta }\,\left|
B_{p^{+}=m+\alpha ,\,p^{-}=-\frac{r}{m+\alpha },\,p_{2}=\rho }\right\rangle .
\label{physB}
\end{equation}
D-branes with this physical boundary state wrap the twisted conjugacy
classes (\ref{twistedD2}) in AdS$_{3}$.

\section{Conclusions}

In this paper we have determined the boundary states for AdS$_{2}$
D-branes in Lorentzian AdS$_{3}$. In constructing the boundary
states we have utilized the free field formalism of SL(2,R)\ WZNW\
model. The gluing conditions among the symmetry currents are
translated into the free field language and the boundary
conditions among the modes of free fields are deduced. Ishibashi
states are constructed as coherent states in the free field Fock
space and the boundary conditions imposed on them are shown to be
satisfied. Annulus amplitude is evaluated in the closed string
channel and then it is
compared to the characters of the unitary representations of $\widehat{%
\mathtt{sl}}$(2,R) Kac-Moody algebra. It is found that only the
closed strings in the principal continuous series of
$\widehat{\mathtt{sl}}$(2,R) couple to AdS$_{2}$ D-branes. This
result is in accordance with the conclusions of
\cite{bars1},\cite{BDM}. Then using the fact that AdS$_{2}$
D-branes are located on the twisted conjugacy classes of SL(2,R),
the form of physical boundary states is determined.

At the beginning of section $4$ we explained the non-usability of
the Cardy procedure in the case of SL(2,R) WZNW model. The method
of finding the boundary states by determining the one-point
functions in BCFT \cite{onepoint},\cite{teschner} is also not
possible in the case of Lorentzian AdS$_3$. In
\cite{onepoint},\cite{teschner} the expressions for one-point
functions are determined by solving the factorization conditions
of two-point functions in Euclidean AdS$_3$. Two-point functions
are known in SL(2,C)/SU(2) WZNW model (Euclidean AdS$_3$), however
the correct procedure to Wick rotate them into SL(2,R) WZNW model
(Lorentzian AdS$_3$) is still not known. Therefore, presently,
utilizing such a technique in Lorentzian AdS$_3$ is not possible.
Only after proper calculation of two-point and three-point
functions in SL(2,R) WZNW model one can use BCFT techniques to
determine the one-point functions and consequently the physical
boundary states. Comparing the physical boundary states obtained
via different methods will be a non-trivial check of the
correctness of their form (\ref{physB}).

As a possible future research these physical boundary states can
be used to calculate the boundary correlation functions. The free
field formalism will help this calculations to simplify
enormously. Such correlation functions correspond to physically
observable quantities and they will provide more information on
the properties of the theory. The regularization and the modular
properties of the annulus amplitude (\ref
{annulus},\ref{holo-amp}) should also be examined thoroughly. From
the regularized and modular transformed form of this amplitude it
should be possible to read the open string spectrum in the
Lorentzian AdS$_{3}$.

\section*{Acknowledgements}

I would like to thank I. Bars for comments on the manuscript. This
research has been supported in part by the Turkish Academy of
Sciences in the framework of the Young Scientist Award Program
(CD/T\"{U}BA--GEB\.{I}P/2002--1--7).



\begin{thebibliography}{99}
\bibitem{BRFW}  J. Balog, L. O'Raighfertaigh, P. Forgacs, and A. Wipf, Nucl.
Phys. \textbf{B325}, 225 (1989).

\bibitem{BN}  I. Bars and D. Nemeschansky, Nucl. Phys. \textbf{B348,} 89
(1991).

\bibitem{others1}  P. M. S. Petropoulos, Phys. Lett. \textbf{B236}, 151
(1990). \newline
N. Mohammedi, Int. J. Mod. Phys. \textbf{A5}, 3201 (1990). \newline
S. Hwang, Nucl. Phys. \textbf{B354}, 100 (1991). \newline
M. Henningson and S. Hwang, Phys. Lett. \textbf{B258}, 341 (1991). \newline
M. Henningson, S. Hwang, P. Roberts, and B. Sundborg, Phys. Lett. \textbf{%
B267}, 350 (1991).

\bibitem{roberts}  S. Hwang and P. Roberts, \textit{``Interaction and
modular invariance of strings on curved manifolds''}, Proceedings of the
16th Johns Hopkins Workshop, \textit{\ Current Problems in Particle Theory}
(Goteborg, Sweden, 1992) [arXiv:hep-th/9211075].

\bibitem{bars1}  I. Bars, Phys. Rev. \textbf{D53}, 3308 (1996)
[arXiv:hep-th/9503205]. \newline
I. Bars, \textit{``Solution of the SL(2,R) String in Curved Spacetime''}, in
Proceedings of Strings'95 Conference, \textit{Future Perspectives in String
Theory}, Eds. I. Bars et al., World Scientific (1996), page 3
[arXiv:hep-th/9511187].

\bibitem{others2}  Y. Satoh, Nucl. Phys. \textbf{B513}, 213 (1998)
[arXiv:hep-th/9705208]. \newline
A. Giveon, D. Kutasov and N. Seiberg,
Adv. Theor. Math. Phys. \textbf{2}, 733 (1998) [arXiv:hep-th/9806194].
\newline
J. de Boer, H. Ooguri, H. Robins and J. Tannenhauser,
JHEP \textbf{9812}, 026 (1998) [arXiv:hep-th/9812046]. \newline
D. Kutasov and N. Seiberg, 
JHEP \textbf{9904}, 008 (1999) [arXiv:hep-th/9903219]. \newline
P. M. S. Petropoulos, \textit{``String theory on $AdS_{3}$: some open
questions''}, Proceedings of TMR European program meeting, \textit{Quantum
Aspects of Gauge Theories, Supersymmetry and Unification} (Paris, France,
1999) [arXiv:hep-th/9908189].

\bibitem{BDM}  I. Bars, C. Deliduman and D. Minic, \textit{''String Theory
on AdS}$_{3}$\textit{\ Revisited''}, [arXiv:hep-th/9907087].

\bibitem{MO-1}  J. M. Maldacena and H. Ooguri,
J. Math. Phys. \textbf{42}, 2929 (2001) [arXiv:hep-th/0001053].\newline
J. M. Maldacena, H.Ooguri and J. Son,
J. Math. Phys. \textbf{42}, 2961 (2001) [arXiv:hep-th/0005183].

\bibitem{witten}  E. Witten, Phys. Rev. \textbf{D44}, 314 (1991).

\bibitem{BTZ}  M. Banados, M. Henneaux, C. Teitelboim and J. Zanelli,
Phys. Rev. \textbf{D48}, 1506 (1993) [arXiv:gr-qc/9302012].

\bibitem{thorn}  C. Thorn, Nucl. Phys.\ \textbf{B248}, 551 (1974).

\bibitem{peskin}  L. J. Dixon, M. E. Peskin and J. Lykken, Nucl. Phys.
\textbf{B325}, 329 (1989).

\bibitem{larsen}  A. L. Larsen and N. Sanchez,
Phys. Rev. \textbf{D62}, 046003 (2000) [arXiv:hep-th/0001180].

\bibitem{TASI}  J. Polchinski, \textit{``TASI lectures on D-branes''},
[arXiv:hep-th/9611050].

\bibitem{stanciu-0}  A. Y. Alekseev and V. Schomerus,
Phys. Rev. \textbf{D60}, 061901 (1999) [arXiv:hep-th/9812193]. \newline
G. Felder, J. Frohlich, J. Fuchs and C. Schweigert,
J. Geom. Phys. \textbf{34}, 162 (2000) [arXiv:hep-th/9909030]. \newline
S. Stanciu, 
JHEP \textbf{0001}, 025 (2000) [arXiv:hep-th/9909163].

\bibitem{stanciu}  S. Stanciu,
JHEP \textbf{9909}, 028 (1999) [arXiv:hep-th/9901122].\newline
J. M. Figueroa-O'Farrill and S. Stanciu,
JHEP \textbf{0004}, 005 (2000) [arXiv:hep-th/0001199].

\bibitem{bachas}  C. Bachas and M. Petropoulos,
JHEP \textbf{0102}, 025 (2001) [arXiv:hep-th/0012234].

\bibitem{lomholt}  M. A. Lomholt and A. L. Larsen, \textit{``Open strings in
the SL(2,R) WZNW model with solution for a rigidly rotating
string''}, [arXiv:hep-th/0107035].

\bibitem{ads2d}  P. M. Petropoulos and S. Ribault,
JHEP {\bf 0107}, 036 (2001) [arXiv:hep-th/0105252].

\bibitem{onepoint} A. Giveon, D. Kutasov and A. Schwimmer,
Nucl. Phys. \textbf{B615}, 133 (2001) [arXiv:hep-th/0106005].
\newline
A. Parnachev and D. A. Sahakyan, 
JHEP \textbf{0110}, 022 (2001) [arXiv:hep-th/0109150]. \newline
P. Lee, H. Ooguri and J. Park,
Nucl. Phys. \textbf{B632}, 283 (2002) [arXiv:hep-th/0112188].

\bibitem{hikida} Y. Hikida and Y. Sugawara,
Prog. Theor. Phys. \textbf{107}, 1245 (2002)
[arXiv:hep-th/0107189].

\bibitem{rajaraman}  A. Rajaraman and M. Rozali,
Phys. Rev. \textbf{D66}, 026006 (2002) [arXiv:hep-th/0108001].

\bibitem{teschner}  B. Ponsot, V. Schomerus and J. Teschner,
JHEP \textbf{0202}, 016 (2002) [arXiv:hep-th/0112198].

\bibitem{ponsot}  S. Ribault, \textit{``AdS$_{2}$ D-branes in AdS$_{3}$
spacetime''}, [arXiv:hep-th/0207094].
\newline
B. Ponsot and S. Silva,
Phys. Lett. {\bf B551}, 173 (2003) [arXiv:hep-th/0209084].
\newline S. Ribault, \textit{``Two AdS(2) branes in the Euclidean
AdS(3)''}, [arXiv:hep-th/0210248].

\bibitem{teschner-1}  J.~Teschner,
Nucl. Phys. \textbf{B546}, 369 (1999) [arXiv:hep-th/9712258].

\bibitem{troost}  J. Troost,
JHEP \textbf{0209}, 041 (2002) [arXiv:hep-th/0206118].

\bibitem{ishikawa}  H. Ishikawa and S. Watamura,
JHEP \textbf{0008}, 044 (2000) [arXiv:hep-th/0007141].

\bibitem{ishibashi1}  N. Ishibashi and T. Onogi,
Nucl. Phys. \textbf{B318}, 239 (1989).

\bibitem{ishibashi2}  N. Ishibashi,
Mod. Phys. Lett. \textbf{A4}, 251 (1989).

\bibitem{cardy}  J. L. Cardy,
Nucl. Phys. \textbf{B324}, 581 (1989).

\bibitem{teschner-0}  J. Teschner, \textit{``Remarks on Liouville Theory
with Boundary''}, PRHEP-tmr2000/041 (Proceedings of TMR-conference, \textit{%
Nonperturbative Quantum Effects 2000}) [arXiv:hep-th/0009138].

\bibitem{kawai}  S. Kawai,
Nucl. Phys. \textbf{B630}, 203 (2002) [arXiv:hep-th/0201146].
\end{thebibliography}
\end{document}